\documentclass[narrowdisplay,onecolumn]{elsart3p}
\usepackage{graphics}
\usepackage{amssymb}
\usepackage{lineno}

\begin{document}
\hfill {\tiny HISKP-TH-07/15, FZJ-IKP-TH-2007-18}

\begin{frontmatter}
\title{Strangeness $S=-2$ baryon-baryon interactions using chiral effective field theory}
\author[1]{H. Polinder\corauthref{cor}},
\corauth[cor]{Corresponding author.}
\ead{h.polinder@fz-juelich.de}
\author[1]{J. Haidenbauer},
\author[1,2]{U.-G. Mei\ss ner}
\address[1]{Institut f{\"u}r Kernphysik (Theorie), Forschungszentrum J{\"u}lich, D-52425 J{\"u}lich, Germany}
\address[2]{Helmholtz-Institut f{\"u}r Strahlen- und Kernphysik (Theorie), Universit{\"a}t Bonn, D-53115 Bonn, Germany}

\begin{abstract}
We derive the leading order strangeness $S=-2$ baryon-baryon interactions in chiral effective field theory. The potential consists of contact terms without derivatives and of one-pseudoscalar-meson exchanges. The contact terms and the couplings of the pseudoscalar mesons to the baryons are related via ${\rm SU(3)}_{\rm f}$ symmetry to the $S=-1$ hyperon-nucleon channels. We show that the chiral effective field theory predictions with natural values for the low-energy constants agree with the experimental information in the $S=-2$ sector.
\end{abstract}
\end{frontmatter}

%%%%%%%%%%%%%%%%%%%%%%%%%%%%%%%%%%%%%%%%%%%%%%%%%%%%%%%%%%%%%%%%%%%%%%%%%%%%%%%

\section{Introduction}
\label{chap:1}
As of today, theoretical investigations of the baryon-baryon interaction in the strangeness $S=-2$ sector were performed within the meson-exchange picture \cite{Stoks:1999bz,Rijken:2006kg,Tominaga:1998iy,Ueda1} as well as in the constituent quark model \cite{Oka:1986fr,Straub:1988mz,Straub1,Koike:1989ak,Oka:1990vx,Nakamoto:1997gh,Fujiwara,Fujiwara:2006yh}. In both approaches the assumption of ${\rm SU(3)}_{\rm f}$ symmetry is an essential prerequisite. It allows to connect the doubly strange cascade-nucleon ($\Xi N$) and hyperon-hyperon ($YY$) interactions ($Y = \Lambda, \Sigma$) to the hyperon-nucleon ($YN$) and nucleon-nucleon ($NN$) interactions, i.e. to systems where a wealth of experimental information is available, which can then be used to constrain the parameters inherent to those approaches. 

Indeed, the experimental knowledge on the $\Xi N$ and $YY$ interactions themselves is quite poor. Until the beginning of this century the only information available came from doubly strange hypernuclei and, moreover, only three candidates for such hypernuclei were reported \cite{Dan63,Prowse:1966nz,Aoki:1991ip}. The $\Lambda\Lambda$ binding energies derived from these events indicated a strongly attractive ${}^1S_0$ $\Lambda\Lambda$ interaction. However, more recently a new candidate for ${}_{\Lambda\Lambda}^{\;\;\;6}{\rm He}$ with a much lower binding energy was identified \cite{Takahashi:2001nm}, the so-called Nagara event, suggesting that the $\Lambda\Lambda$ interaction should be only moderately attractive. This conjecture is also in line with evidence provided by the latest searches for the $H$ dibaryon, a bound state in the $S=-2$ sector proposed by Jaffe back in 1977 \cite{Jaffe:1976yi}, whose existence is now considered to be practically ruled out \cite{Yoon:2007aq}. (See also the theoretical works \cite{Reuber,Oset} concerning the attraction in the $\Lambda\Lambda$ system.) Very recently doubly strange baryon-baryon scattering data at lower energies, below $p_{\rm lab}=0.8$ GeV, were deduced for the first time \cite{Tamagawa:2001tk,Ahn:2005jz}. An upper limit of $24$ mb at $90\%$ confidence level was provided for elastic $\Xi^-p$ scattering, and for the $\Xi^-p\rightarrow \Lambda\Lambda$ cross section at $p_{\rm lab}=500$ MeV a value of $4.3^{+6.3}_{-2.7}$ mb was reported \cite{Ahn:2005jz}.

Over the last decade a new powerful tool for understanding hadronic interactions has emerged, namely {\em chiral effective field theory} (EFT). This approach, which was pioneered by Weinberg in the early nineties, incorporates explicitly the scales and symmetries of Quantum Chromodynamics. An important advantage of EFT is that there is an underlying power counting that allows to improve calculations systematically by going to higher orders in a perturbative expansion and, at the same time, it allows to estimate theoretical uncertainties. In addition, it is possible to derive two- and corresponding three-baryon forces in a consistent way. The concepts of chiral EFT have been applied in the last decade to the $NN$ interaction and to the physics of light nuclei, resulting in a high-precision description of the experimental data, see e.g. Refs.~\cite{Bedaque:2002mn,Epelbaum:2005pn} and references therein. Recently we utilized the chiral EFT framework for investigating the $YN$ interaction. In particular, we showed that the leading order (LO) chiral EFT successfully describes the available $YN$ scattering data \cite{Polinder:2006zh}. Also the binding energies of the light hypernuclei are predicted well within chiral EFT \cite{Nogga:2006ir,Haidenbauer:2007ra}. 

The $\Xi N$ and $YY$ interactions have not been studied using chiral EFT so far. In this letter we report on the first chiral EFT investigation of the $S=-2$ sector, starting with a LO calculation. For this purpose we extend the recently constructed LO chiral EFT potential of the $YN$ interaction \cite{Polinder:2006zh}. We employ ${\rm SU(3)}_{\rm f}$ relations to connect the doubly strange with the singly strange channels and we confront the LO chiral EFT predictions with the (poor) experimental knowledge.

%%%%%%%%%%%%%%%%%%%%%%%%%%%%%%%%%%%%%%%%%%%%%%%%%%%%%%%%%%%%%%%%%%%%%%%%%%%%%%%

\section{The effective strangeness $S=-2$ baryon-baryon potential}
\label{chap:2}
We construct the chiral effective potentials for the $S=-2$ sector at LO using the Weinberg power counting similar to the $YN$ case considered in \cite{Polinder:2006zh}. The LO potential consists of four-baryon contact terms without derivatives and of one-pseudoscalar-meson exchanges. The LO ${\rm SU(3)}_{\rm f}$ invariant contact terms for the octet baryon-baryon interactions that are Hermitian and invariant under Lorentz transformations were discussed in \cite{Polinder:2006zh,Haidenbauer:2007ra} and we refer the reader to these works for details. The pertinent Lagrangians read
\begin{eqnarray}
{\mathcal L}^1 &=& C^1_i \left<\bar{B}_a\bar{B}_b\left(\Gamma_i B\right)_b\left(\Gamma_i B\right)_a\right>\ , \quad
{\mathcal L}^2 = C^2_i \left<\bar{B}_a\left(\Gamma_i B\right)_a\bar{B}_b\left(\Gamma_i B\right)_b\right>\ , \nonumber \\
{\mathcal L}^3 &=& C^3_i \left<\bar{B}_a\left(\Gamma_i B\right)_a\right>\left<\bar{B}_b\left(\Gamma_i B\right)_b\right>\  .
\label{eq:2.1}
\end{eqnarray}
Here, the labels $a$ and $b$ are the Dirac indices of the particles, the label $i$ denotes the five elements of the Clifford algebra, $B$ is the usual irreducible octet representation of ${\rm SU(3)}_{\rm f}$ given by
\begin{eqnarray}
B&=&
\left(
\begin{array}{ccc}
\frac{\Sigma^0}{\sqrt{2}}+\frac{\Lambda}{\sqrt{6}} & \Sigma^+ & p \\
\Sigma^- & \frac{-\Sigma^0}{\sqrt{2}}+\frac{\Lambda}{\sqrt{6}} & n \\
-\Xi^- & \Xi^0 & -\frac{2\Lambda}{\sqrt{6}}
\end{array}
\right) \ ,
\label{eq:2.2}
\end{eqnarray}
and the brackets denote taking the trace in the three-dimensional flavor space. The Clifford algebra elements are here actually diagonal $3\times 3$-matrices.

The $\Xi N$ and $YY$ partial wave potentials derived from the above Lagrangians are given in Table~\ref{tab:2.1} for the singlet S-waves and in Table~\ref{tab:2.1a} for the triplet S-waves. The coefficients $C_S$ and $C_T$ are linear combinations of the low-energy coefficients $C_i$'s in Eq.~(\ref{eq:2.1}) and refer to the central and spin-spin parts of the potential, see e.g. Ref. \cite{Polinder:2006zh}. The $S=0$ and $-1$ potentials are listed also in Tables~\ref{tab:2.1} and \ref{tab:2.1a} for completeness. Using the ${\rm SU(3)}_{\rm f}$ Clebsch-Gordan coefficients one can express the baryon-baryon potentials in terms of the ${\rm SU(3)}_{\rm f}$ irreducible representations, see e.g. \cite{deSwart:1963gc} and also \cite{Dover:1991sh}. The last columns of Tables~\ref{tab:2.1} 
and \ref{tab:2.1a} show the ${\rm SU(3)}_{\rm f}$ content of the various potentials.
\begin{table}[t]
\caption{Various LO baryon-baryon contact potentials for the ${}^1S_0$ partial wave in isospin basis. These potentials are flavor symmetric. $V^{27}$ etc. refers to the corresponding ${\rm SU(3)}_{\rm f}$ irreducible representation. 
}
\label{tab:2.1}
\vspace{0.2cm}
\centering
\begin{tabular}{|l|c|c|l|l|}
\hline
&Channel &Isospin &$V_{1S0}$&${\rm SU(3)}_{\rm f}$ content\\
\hline
$S=0$&$NN\rightarrow NN$ &$1$ &$4\pi\left[2\left(C^2_S-3C^2_T\right)+2\left(C^3_S-3C^3_T\right) \right]$ & $V^{27}$\\
\hline
$S=-1$&$\Lambda N \rightarrow \Lambda N$ &$\frac{1}{2}$ &$4\pi\left[\frac{1}{6}\left(C^1_S-3C^1_T\right)+\frac{5}{3}\left(C^2_S-3C^2_T\right)+2\left(C^3_S-3C^3_T\right) \right]$ &$\frac{1}{10}\left(9V^{27}+V^{8_s}\right)$\\
&$\Lambda N \rightarrow \Sigma N$ &$\frac{1}{2}$ &$4\pi\left[\frac{1}{2}\left(C^1_S-3C^1_T\right)-\left(C^2_S-3C^2_T\right) \right]$ &$\frac{3}{10}\left(-V^{27}+V^{8_s}\right)$\\
&$\Sigma N \rightarrow \Sigma N$ &$\frac{1}{2}$ &$4\pi\left[\frac{3}{2}\left(C^1_S-3C^1_T\right)-\left(C^2_S-3C^2_T\right)+2\left(C^3_S-3C^3_T\right) \right]$ &$\frac{1}{10}\left(V^{27}+9V^{8_s}\right)$\\
&$\Sigma N \rightarrow \Sigma N$ &$\frac{3}{2}$ &$4\pi\left[2\left(C^2_S-3C^2_T\right)+2\left(C^3_S-3C^3_T\right) \right]$ &$V^{27}$\\
\hline
$S=-2$&$\Lambda\Lambda \rightarrow \Lambda\Lambda$ &$0$ &$4\pi\left[\left(C^1_S-3C^1_T\right)+\left(C^2_S-3C^2_T\right)+2\left(C^3_S-3C^3_T\right) \right]$ & $\frac{1}{40}\left(27V^{27}+8V^{8_s}+5V^{1}\right)$\\
&$\Lambda\Lambda \rightarrow \Xi N$ &$0$ &$4\pi\left[\frac{5}{3}\left(C^1_S-3C^1_T\right)-\frac{4}{3}\left(C^2_S-3C^2_T\right) \right]$ &$\frac{-1}{40}\left(18V^{27}-8V^{8_s}-10V^{1}\right)$\\
&$\Lambda\Lambda \rightarrow \Sigma\Sigma$ &$0$ &$4\pi\left[-\frac{\sqrt{3}}{3}\left(C^1_S-3C^1_T\right)-\frac{\sqrt{3}}{3}\left(C^2_S-3C^2_T\right) \right]$ &$\frac{\sqrt{3}}{40}\left(-3V^{27}+8V^{8_s}-5V^{1}\right)$\\
&$\Xi N \rightarrow \Xi N$ &$0$ &$4\pi\left[3\left(C^1_S-3C^1_T\right)+2\left(C^3_S-3C^3_T\right) \right]$ &$\frac{1}{40}\left(12V^{27}+8V^{8_s}+20V^{1}\right)$\\
&$\Xi N \rightarrow \Sigma\Sigma$ &$0$ &$4\pi\left[-\sqrt{3}\left(C^1_S-3C^1_T\right) \right]$ &$\frac{\sqrt{3}}{40}\left(2V^{27}+8V^{8_s}-10V^{1}\right)$\\
&$\Sigma\Sigma \rightarrow \Sigma\Sigma$ &$0$ &$4\pi\left[3\left(C^1_S-3C^1_T\right)-\left(C^2_S-3C^2_T\right)+2\left(C^3_S-3C^3_T\right) \right]$ &$\frac{1}{40}\left(V^{27}+24V^{8_s}+15V^{1}\right)$\\
%\hline
&$\Xi N \rightarrow \Xi N$ &$1$ &$4\pi\left[\left(C^1_S-3C^1_T\right)+2\left(C^3_S-3C^3_T\right) \right]$ &$\frac{1}{5}\left(2V^{27}+3V^{8_s}\right)$\\
&$\Xi N \rightarrow \Sigma\Lambda$ &$1$ &$4\pi\left[-\frac{\sqrt{6}}{3}\left(C^1_S-3C^1_T\right)+\frac{2\sqrt{6}}{3}\left(C^2_S-3C^2_T\right) \right]$ &$\frac{\sqrt{6}}{5}\left(V^{27}-V^{8_s}\right)$\\
&$\Sigma\Lambda \rightarrow \Sigma\Lambda$ &$1$ &$4\pi\left[\frac{2}{3}\left(C^1_S-3C^1_T\right)+\frac{2}{3}\left(C^2_S-3C^2_T\right)+2\left(C^3_S-3C^3_T\right) \right]$ &$\frac{1}{5}\left(3V^{27}+2V^{8_s}\right)$\\
%\hline
&$\Sigma\Sigma \rightarrow \Sigma\Sigma$ &$2$ &$4\pi\left[2\left(C^2_S-3C^2_T\right)+2\left(C^3_S-3C^3_T\right) \right]$ &$V^{27}$\\
\hline
\end{tabular}
\end{table}
\begin{table}[t]
\caption{Various LO baryon-baryon contact potentials for the ${}^3S_1$ partial wave in the isospin basis. These potentials are flavor antisymmetric. $V^{10^*}$ etc. refers to the corresponding ${\rm SU(3)}_{\rm f}$ irreducible representation. 
}
\label{tab:2.1a}
\vspace{0.2cm}
\centering
\begin{tabular}{|l|c|c|l|l|}
\hline
&Channel &Isospin &$V_{3S1}$&${\rm SU(3)}_{\rm f}$ content\\
\hline
$S=0$&$NN \rightarrow NN$ &$1$ &$4\pi\left[2\left(C^2_S+C^2_T\right)+2\left(C^3_S+C^3_T\right) \right]$ &$V^{10^*}$\\
\hline
$S=-1$&$\Lambda N \rightarrow \Lambda N$ &$\frac{1}{2}$ &$4\pi\left[\frac{3}{2}\left(C^1_S+C^1_T\right)+\left(C^2_S+C^2_T\right)+2\left(C^3_S+C^3_T\right) \right]$ &$\frac{1}{2}\left(V^{8_a}+V^{10^*}\right)$\\
&$\Lambda N \rightarrow \Sigma N$ &$\frac{1}{2}$ &$4\pi\left[-\frac{3}{2}\left(C^1_S+C^1_T\right)+\left(C^2_S+C^2_T\right) \right]$ &$\frac{1}{2}\left(-V^{8_a}+V^{10^*}\right)$\\
&$\Sigma N \rightarrow \Sigma N$ &$\frac{1}{2}$ &$4\pi\left[\frac{3}{2}\left(C^1_S+C^1_T\right)+\left(C^2_S+C^2_T\right)+2\left(C^3_S+C^3_T\right) \right]$ &$\frac{1}{2}\left(V^{8_a}+V^{10^*}\right)$\\
&$\Sigma N \rightarrow \Sigma N$ &$\frac{3}{2}$ &$4\pi\left[-2\left(C^2_S+C^2_T\right)+2\left(C^3_S+C^3_T\right) \right]$ &$V^{10}$\\
\hline
$S=-2$&$\Xi N \rightarrow \Xi N$ &$0$ &$4\pi\left[3\left(C^1_S+C^1_T\right)+2\left(C^3_S+C^3_T\right) \right]$ &$V^{8_a}$\\
%\hline
&$\Xi N \rightarrow \Xi N$ &$1$ &$4\pi\left[\left(C^1_S+C^1_T\right)+2\left(C^3_S+C^3_T\right) \right]$ &$\frac{1}{3}\left(V^{10}+V^{10^*}+V^{8_a}\right)$\\
&$\Xi N \rightarrow \Sigma\Lambda$ &$1$ &$4\pi\left[-\frac{2}{3}\sqrt{6}\left(C^2_S+C^2_T\right) \right]$ &$\frac{\sqrt{6}}{6}\left(V^{10}-V^{10^*}\right)$\\
&$\Xi N \rightarrow \Sigma\Sigma$ &$1$ &$4\pi\left[-\sqrt{2}\left(C^1_S+C^1_T\right) \right]$ &$\frac{\sqrt{2}}{6}\left(V^{10}+V^{10^*}-2V^{8_a}\right)$\\
&$\Sigma\Lambda \rightarrow \Sigma\Lambda$ &$1$ &$4\pi\left[2\left(C^3_S+C^3_T\right) \right]$ &$\frac{1}{2}\left(V^{10}+V^{10^*}\right)$\\
&$\Sigma\Lambda \rightarrow \Sigma\Sigma$ &$1$ &$4\pi\left[-\frac{2}{3}\sqrt{3}\left(C^2_S+C^2_T\right) \right]$ &$\frac{\sqrt{3}}{6}\left(V^{10}-V^{10^*}\right)$\\
&$\Sigma\Sigma \rightarrow \Sigma\Sigma$ &$1$ &$4\pi\left[2\left(C^1_S+C^1_T\right)+2\left(C^3_S+C^3_T\right) \right]$ &$\frac{1}{6}\left(V^{10}+V^{10^*}+4V^{8_a}\right)$\\
\hline
\end{tabular}
\end{table}
Contrary to the Nijmegen meson-exchange models \cite{Stoks:1999bz,Rijken:2006kg} or the constituent quark model \cite{Fujiwara:2006yh}, say, the chiral EFT for the $S=-2$ channels is not completely fixed by imposing ${\rm SU(3)}_{\rm f}$ symmetry. The symmetry connects only five of the six LO contact terms to those appearing in the $NN$ and $YN$ interactions. Thus, one contact term remains undetermined. In the present study those five LO contact terms are taken over from Ref.~\cite{Polinder:2006zh}, where they were fixed by a fit to the $YN$ data. The additional contact term, which occurs in the singlet isospin-zero doubly strange $\Xi N$ and $YY$ channels, needs to be determined, in principle, from empirical information on the $S=-2$ baryon-baryon interaction. For the additional contact term we have chosen the projection on the $\Lambda\Lambda$ singlet S-wave: $C^{\Lambda\Lambda,\Lambda\Lambda}_{1S0}$. All other LO $\Xi N$ and $YY$ contact potentials are then fixed by ${\rm SU(3)}_{\rm f}$ symmetry.

%\subsection{One pseudoscalar-meson exchange}
%\label{chap:2.2}
The lowest order ${\rm SU(3)}_{\rm f}$ invariant pseudoscalar-meson--baryon
interaction Lagrangian with the appropriate symmetries was discussed in \cite{Polinder:2006zh}. In the isospin basis it reads
\begin{eqnarray}
{\mathcal L}&=&-f_{NN\pi}\bar{N}\gamma^\mu\gamma_5\mbox{\boldmath $\tau$}N\cdot\partial_\mu\mbox{\boldmath $\pi$} +if_{\Sigma\Sigma\pi}\bar{\mbox{\boldmath $ \Sigma$}}\gamma^\mu\gamma_5\times{\mbox{\boldmath $ \Sigma$}}\cdot\partial_\mu\mbox{\boldmath $\pi$} \nonumber \\
&&-f_{\Lambda\Sigma\pi}\left[\bar{\Lambda}\gamma^\mu\gamma_5{\mbox{\boldmath $ \Sigma$}}+\bar{\mbox{\boldmath $\Sigma$}}\gamma^\mu\gamma_5\Lambda\right]\cdot\partial_\mu\mbox{\boldmath $\pi$}-f_{\Xi\Xi\pi}\bar{\Xi}\gamma^\mu\gamma_5\mbox{\boldmath $\tau$}\Xi\cdot\partial_\mu\mbox{\boldmath $\pi$} \nonumber \\
&&-f_{\Lambda NK}\left[\bar{N}\gamma^\mu\gamma_5\Lambda\partial_\mu K+\bar{\Lambda}\gamma^\mu\gamma_5N\partial_\mu K^\dagger\right]
\nonumber \\&&
-f_{\Xi\Lambda K}\left[\bar{\Xi}\gamma^\mu\gamma_5\Lambda\partial_\mu K_c+\bar{\Lambda}\gamma^\mu\gamma_5\Xi\partial_\mu K_c^\dagger\right]
\nonumber \\&&
-f_{\Sigma NK}\left[\bar{\mbox{\boldmath $ \Sigma$}}\cdot\gamma^\mu\gamma_5\partial_\mu K^\dagger\mbox{\boldmath $\tau$}N+\bar{N}\gamma^\mu\gamma_5\mbox{\boldmath $\tau$}\partial_\mu K\cdot{\mbox{\boldmath $ \Sigma$}}\right]
\nonumber \\&&
-f_{\Xi \Sigma K}\left[\bar{\mbox{\boldmath $ \Sigma$}}\cdot\gamma^\mu\gamma_5\partial_\mu K_c^\dagger\mbox{\boldmath $\tau$}\Xi+\bar{\Xi}\gamma^\mu\gamma_5\mbox{\boldmath $\tau$}\partial_\mu K_c\cdot{\mbox{\boldmath $ \Sigma$}}\right]
%\nonumber \\&&
-f_{NN\eta_8}\bar{N}\gamma^\mu\gamma_5N\partial_\mu\eta
\nonumber \\&&
-f_{\Lambda\Lambda\eta_8}\bar{\Lambda}\gamma^\mu\gamma_5\Lambda\partial_\mu\eta-f_{\Sigma\Sigma\eta_8}\bar{\mbox{\boldmath $ \Sigma$}}\cdot\gamma^\mu\gamma_5{\mbox{\boldmath $ \Sigma$}}\partial_\mu\eta
-f_{\Xi\Xi\eta_8}\bar{\Xi}\gamma^\mu\gamma_5\Xi\partial_\mu\eta \ .
\label{eq:2.3}
\end{eqnarray}
%We have introduced the isospin doublets
%\begin{equation}
%N=\left(\begin{array}{r}p\\n\end{array}\right)\ ,\ \ \Xi=\left(\begin{array}{r}\Xi^0\\\Xi^-\end{array}\right)\ ,\ \ K=\left(\begin{array}{r}K^+\\K^0\end{array}\right)\ ,\ \ K_c=\left(\begin{array}{r}\bar{K}^0\\-K^-\end{array}\right)\ .
%\label{eq:2.4}
%\end{equation}
The interaction Lagrangian in Eq.~(\ref{eq:2.3}) is invariant under ${\rm SU(3)}_{\rm f}$ transformations if the various coupling constants fulfill specific relations which can be expressed in terms of the coupling constant $f$ and the $F/(F+D)$-ratio $\alpha$ as \cite{deSwart:1963gc},
\begin{equation}
\begin{array}{rlrlrl}
f_{NN\pi}  = & f, & f_{NN\eta_8}  = & \frac{1}{\sqrt{3}}(4\alpha -1)f, & f_{\Lambda NK} = & -\frac{1}{\sqrt{3}}(1+2\alpha)f, \\
f_{\Xi\Xi\pi}  = & -(1-2\alpha)f, &  f_{\Xi\Xi\eta_8}  = & -\frac{1}{\sqrt{3}}(1+2\alpha )f, & f_{\Xi\Lambda K} = & \frac{1}{\sqrt{3}}(4\alpha-1)f, \\
f_{\Lambda\Sigma\pi}  = & \frac{2}{\sqrt{3}}(1-\alpha)f, & f_{\Sigma\Sigma\eta_8}  = & \frac{2}{\sqrt{3}}(1-\alpha )f, & f_{\Sigma NK} = & (1-2\alpha)f, \\
f_{\Sigma\Sigma\pi}  = & 2\alpha f, &  f_{\Lambda\Lambda\eta_8}  = & -\frac{2}{\sqrt{3}}(1-\alpha )f, & f_{\Xi\Sigma K} = & -f.
\end{array}
\label{eq:2.5}
\end{equation}
Here $f\equiv g_A/2F_\pi$, $g_A$ is the axial-vector strength, $g_A= 1.26$, which is measured in neutron $\beta$--decay and $F_\pi$ is the weak pion decay constant, $F_\pi =  92.4$ MeV. For the $F/(F+D)$-ratio we adopt here the SU(6) value ($\alpha=0.4$) which was already used in our study of the $YN$ system \cite{Polinder:2006zh}. The spin-space part of the one-pseudoscalar-meson-exchange potential resulting from the interaction Lagrangian Eq. (\ref{eq:2.3}) is in leading order similar to the static one-pion-exchange potential in \cite{Epelbaum:1998ka},
\begin{eqnarray}
V^{B_1B_2\to B_1'B'_2}&=&-f_{B_1B'_1P}f_{B_2B'_2P}\frac{\left(\mbox{\boldmath $\sigma$}_1\cdot{\bf k}\right)\left(\mbox{\boldmath $
\sigma$}_2\cdot{\bf k}\right)}{{\bf k}^2+m^2_P}\ ,
\label{eq:2.6}
\end{eqnarray}
where $f_{B_1B'_1P}$, $f_{B_2B'_2P}$ are the appropriate coupling constants as given in Eq. (\ref{eq:2.5}) and $m_P$ is the actual mass of the exchanged pseudoscalar meson. With regard to the $\eta$ meson we identified its coupling with the octet value, i.e. the one for $\eta_8$. We defined the transferred and average momentum, ${\bf k}$ and ${\bf q}$, in terms of the final and initial center-of-mass (c.m.) momenta of the baryons, ${\bf p}_f$ and ${\bf p}_i$, as ${\bf k}={\bf p}_f-{\bf p}_i$ and ${\bf q}=({\bf p}_f+{\bf p}_i)/2$. To find the complete LO one-pseudoscalar-meson-exchange potential one needs to multiply the potential in Eq.~(\ref{eq:2.6}) with the isospin factors given in Table~\ref{tab:2.2}.
\begin{table}[t]
%\begin{table}[h]
\caption{The isospin factors for the various one-pseudoscalar-meson exchanges. $P_f$ is the flavor-exchange operator, a dash denotes a non-existing channel.}
\label{tab:2.2}
\vspace{0.2cm}
\centering
\begin{tabular}{|c|c|c|c|c|}
%\begin{tabular}{rrrrr}
\hline
Channel &Exchange &$I=0$ &$I=1$ &$I=2$\\
\hline
$\Lambda\Lambda \rightarrow \Lambda\Lambda$ &$\eta$ &$\frac{1}{2}\left(1+P_f\right)$  &--- &--- \\       
$\Xi N \rightarrow \Xi N$ &$\eta$ &$1$ &$1$ &--- \\
$\Sigma\Sigma \rightarrow \Sigma\Sigma$ &$\eta$ &$\frac{1}{2}\left(1+P_f\right)$ &$\frac{1}{2}\left(1-P_f\right)$ &$\frac{1}{2}\left(1+P_f\right)$ \\
$\Sigma\Lambda \rightarrow \Sigma\Lambda$ &$\eta$ &--- &$1$ &--- \\
$\Xi N \rightarrow \Xi N$ &$\pi$ &$-3$ &$1$ &--- \\
$\Sigma\Sigma \rightarrow \Sigma\Sigma$ &$\pi$ &$-\left(1+P_f\right)$ &$-\frac{1}{2}\left(1-P_f\right)$ &$\frac{1}{2}\left(1+P_f\right)$ \\
$\Lambda\Lambda \rightarrow \Sigma\Sigma$ &$\pi$ &$-\frac{\sqrt{3}}{2}\left(1+P_f\right)$ &--- &--- \\
$\Sigma\Lambda \rightarrow \Lambda\Sigma$ &$\pi$ &--- &$P_f$ &--- \\
$\Sigma\Sigma \rightarrow \Sigma\Lambda$ &$\pi$ &--- &$1-P_f$ &--- \\
$\Lambda\Lambda \rightarrow \Xi N$ &$K$ &$1+P_f$ &--- &--- \\
$\Sigma\Sigma \rightarrow \Xi N$ &$K$ &$\sqrt{3}\left(1+P_f\right)$ &$\sqrt{2}\left(1-P_f\right)$ &--- \\
$\Xi N \rightarrow \Sigma\Lambda$ &$K$ &--- &$\sqrt{2}$ &--- \\
$\Xi N \rightarrow \Lambda\Sigma$ &$K$ &--- &$-\sqrt{2}P_f$ &--- \\
\hline
\end{tabular}
\end{table}
We want to remark that for the $\Xi N$ and $YY$ interactions couplings between channels with non-identical and with identical particles occur which requires special attention \cite{Miyagawa:1997ka}. We follow the treatment of the flavor-exchange potentials as done by the Nijmegen group. Then the proper anti-symmetrization of the states is achieved by multiplying specific transitions with $\sqrt{2}$ factors that are included in Table~\ref{tab:2.2}, see Refs. \cite{Stoks:1999bz,Rijken:2006kg}. In Table~\ref{tab:2.2} $P_f$ is the flavor-exchange operator having the values $P_f=1$ for even-L singlet and odd-L triplet partial waves (antisymmetric in spin-space), and $P_f=-1$ for odd-L singlet and even-L triplet partial waves (symmetric in spin-space). We note that for $\Lambda\Lambda\rightarrow\Lambda\Lambda$, for example, $\eta$-exchange contributes only to spin-space antisymmetric i.e. flavor symmetric partial waves, i.e. ${}^1S_0$,${}^3P_{0,1,2}$, etc.

The ${\rm SU(3)}_{\rm f}$ symmetry of the one-pseudoscalar-meson exchanges is broken by the masses of the pseudoscalar mesons. This is taken into account explicitly in Eq.~(\ref{eq:2.6}) by taking the appropriate values for $m_P$. In case one would consider identical pseudoscalar-meson masses, the corresponding potential obeys the ${\rm SU(3)}_{\rm f}$ relations as shown in the last column of Tables~\ref{tab:2.1} and \ref{tab:2.1a}. This can easily be checked by assuming equal masses and adding the contributions of all one-pseudoscalar-meson exchanges for each channel -- using Eqs.~(\ref{eq:2.5}), (\ref{eq:2.6}) and Table~\ref{tab:2.2} -- and compare the result with the last column of Tables~\ref{tab:2.1} and \ref{tab:2.1a}.

Finally, for completeness we briefly comment on the used scattering equation. The calculations are done in momentum space. We solve the coupled channels (nonrelativistic) Lippmann-Schwinger (LS) equation,
\begin{eqnarray}
T_{\rho''\rho'}^{\nu''\nu',J}(p'',p';\sqrt{s})&=&V_{\rho''\rho'}^{\nu''\nu',J}(p'',p')+
% \nonumber\\&&
\sum_{\rho,\nu}\int_0^\infty \frac{dpp^2}{(2\pi)^3} \, V_{\rho''\rho}^{\nu''\nu,J}(p'',p) 
\frac{2\mu_{\nu}}{q_{\nu}^2-p^2+i\eta}T_{\rho\rho'}^{\nu\nu',J}(p,p';\sqrt{s})\ .
%\nonumber
\end{eqnarray}
The label $\nu$ indicates the particle channels and the label $\rho$ the partial wave. $\mu_\nu$ is the pertinent reduced mass. The on-shell momentum in the intermediate state, $q_{\nu}$, is defined by $\sqrt{s}=\sqrt{M^2_{B_{1,\nu}}+q_{\nu}^2}+\sqrt{M^2_{B_{2,\nu}}+q_{\nu}^2}$. Relativistic kinematics is used for relating the laboratory energy $T_{{\rm lab}}$ of the hyperons to the c.m. momentum. Suppressing the particle channels label, the partial wave projected potentials $V_{\rho''\rho'}^{J}(p'',p')$ are given in \cite{Polinder:2006zh}. The LS
equation for the $\Xi N$ and $YY$ systems is solved in the particle basis, in order to incorporate the correct physical thresholds. The potential in the LS equation is cut off with the regulator function $f^\Lambda(p',p)$,
\begin{equation}
f^\Lambda(p',p)=e^{-\left(p'^4+p^4\right)/\Lambda^4}\ ,
\end{equation}
in order to remove high-energy components of the baryon and pseudoscalar meson fields. The cross sections are calculated using the (LSJ basis) partial wave amplitudes, for details we refer to \cite{Nag75,Holzenkamp:1989tq}. 

%%%%%%%%%%%%%%%%%%%%%%%%%%%%%%%%%%%%%%%%%%%%%%%%%%%%%%%%%%%%%%%%%%%%%%%%%%%%%%%

\section{Results and discussion}
\label{chap:4}
The LO chiral EFT interaction for the $S=-2$ baryon-baryon sector depends, in principle, on six LO contact terms. Five of those contact terms enter also in the $YN$ interaction. Thus, for those we can take over the values which were fixed in our study of the $YN$ sector \cite{Polinder:2006zh}. Indeed only the $\Xi N$ and $YY$ channels that contain the isospin-zero interaction depend on the sixth contact term, which is not fixed yet. The interaction in the other $S=-2$ channels are genuine predictions that follow from the results of Ref.~\cite{Polinder:2006zh} and ${\rm SU(3)}_{\rm f}$ symmetry.

The experimental knowledge on the doubly strange baryon-baryon interaction is quite poor, but since the observation of the Nagara event \cite{Takahashi:2001nm} it is generally accepted that the $\Lambda\Lambda$ interaction is only moderately attractive. Also, very recently an inelastic $\Xi^-p\rightarrow\Lambda\Lambda$ cross section has been deduced at a laboratory momentum of 500 MeV and, in addition, an upper limit with 90\% confidence level has been provided for the elastic $\Xi^-p\rightarrow\Xi^-p$ cross section for laboratory momenta in the range of 200-800 MeV, see Ref.~\cite{Ahn:2005jz}.

Since one contact terms is not yet fixed, we investigate whether those mentioned experimental scattering cross sections of the baryon-baryon interaction in the $S=-2$ sector constrain this additional contact term, $C^{\Lambda\Lambda,\Lambda\Lambda}_{1S0}$. For this purpose we evaluate the relevant doubly strange baryon-baryon cross sections and study their dependence on this additional contact term. This is done for a fixed cut-off value, namely
$\Lambda$ = 600 MeV. 
\begin{figure}[t]
\resizebox{\textwidth}{!}{%
  \includegraphics*[2.0cm,17.0cm][15.65cm,26.8cm]{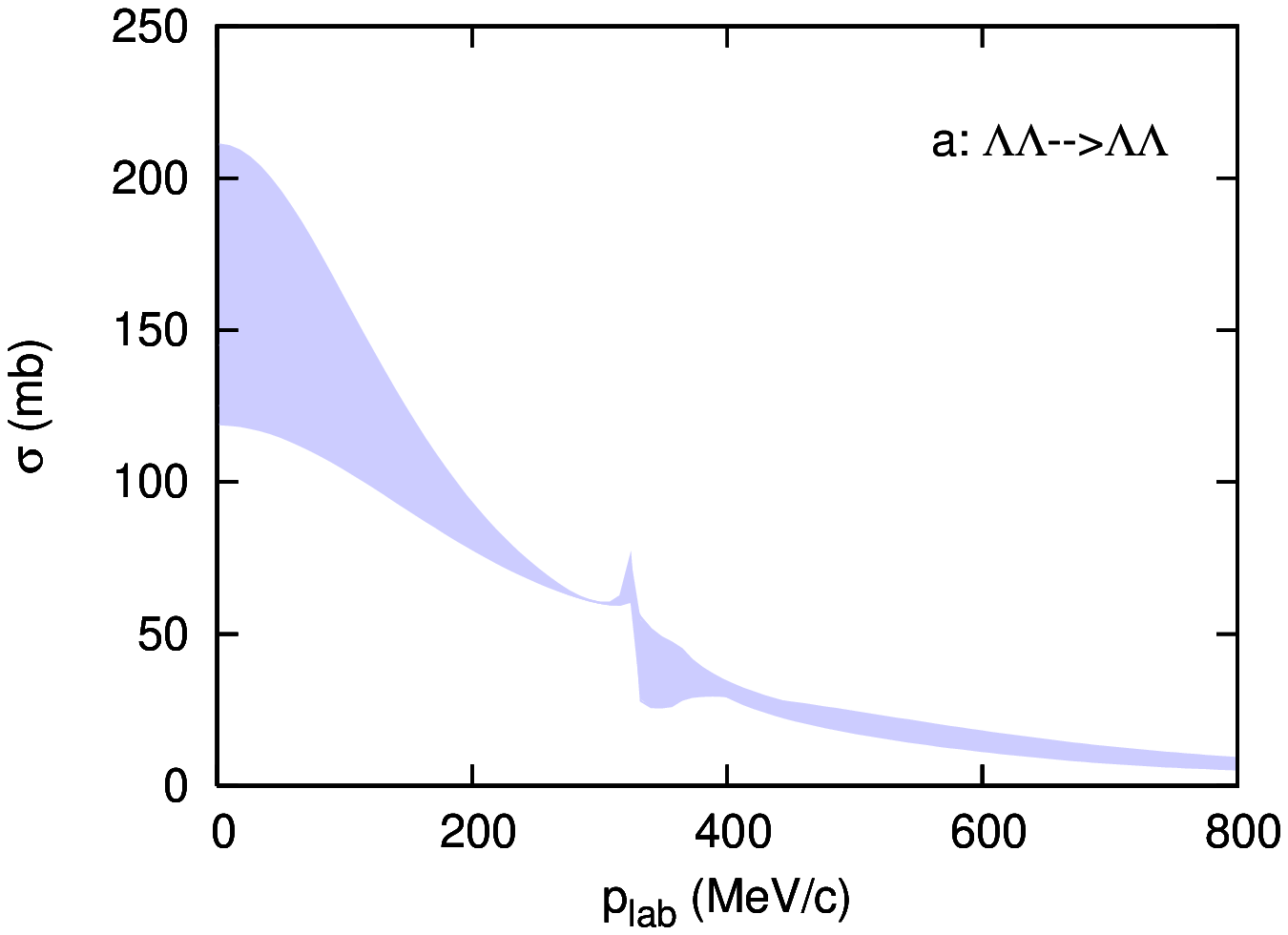}
  \includegraphics*[2.0cm,17.0cm][15.65cm,26.8cm]{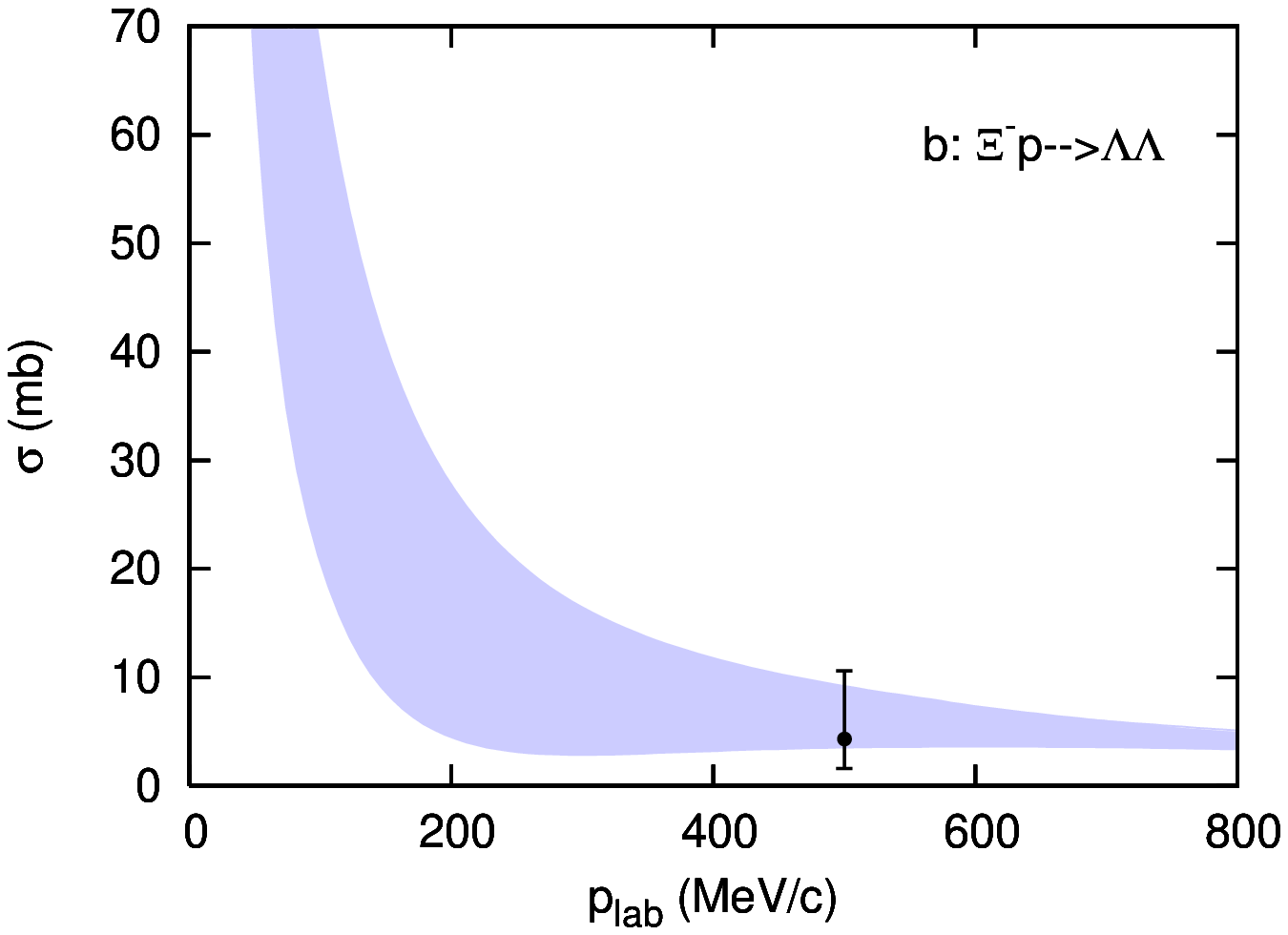}
}
\hfill \break
\resizebox{0.5\textwidth}{!}{%
  \includegraphics*[2.0cm,17.0cm][15.65cm,26.8cm]{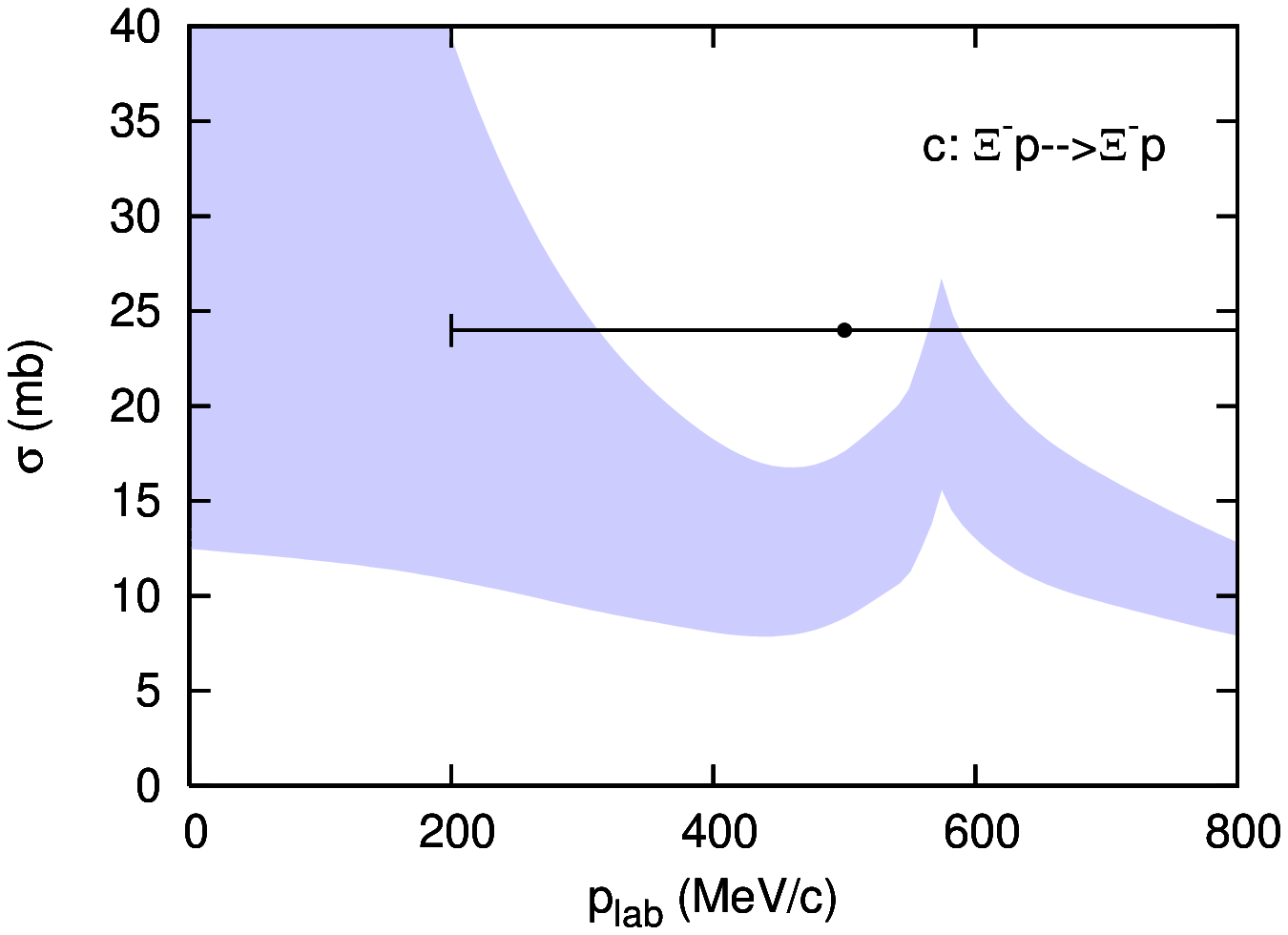}
}
\caption{Total cross sections for the reactions $\Lambda\Lambda \to \Lambda\Lambda$, $\Xi^-p \to \Lambda\Lambda$, and $\Xi^-p \to \Xi^-p$ as a function of $p_{\rm lab}$. The shaded band shows the chiral EFT predictions for variations of the additional contact term $C^{\Lambda\Lambda,\Lambda\Lambda}_{1S0}$ as discussed in the text. The experimental cross sections are taken from Ref.~\cite{Ahn:2005jz}}
\label{fig:4.1}
\end{figure}

We restricted the variations of $|C^{\Lambda\Lambda,\Lambda\Lambda}_{1S0}|$ 
to a range less then twice the natural value, which is equal to $4\pi/f_{\pi}^2$ for this partial-wave projected contact term \cite{Epelbaum:2005pn}. 
Since the $\Lambda\Lambda$ interaction is expected to be only moderately attractive, as mentioned, we considered only such variations of $C^{\Lambda\Lambda,\Lambda\Lambda}_{1S0}$ where the absolute value of the resulting $\Lambda\Lambda$ scattering length was less than $2$ fm. For the same reason, we excluded regions which led to bound states or near-threshold resonances in the $\Lambda\Lambda$ system, which are very unlikely to exist in view of the available experimental information. Based on these considerations the additional contact term $C^{\Lambda\Lambda,\Lambda\Lambda}_{1S0}$ was varied in the range 2.0,...,-0.05 times the natural value. The corresponding results are depicted by the bands in Fig.~\ref{fig:4.1}. (Note that we show only the purely hadronic cross sections. The Coulomb interaction is not taken into account in the present study.) From Figs.~\ref{fig:4.1} (b) and (c) we conclude that the chiral EFT results are consistent with the recently deduced scattering cross sections in the elastic and inelastic $\Xi^-p$ channels. But it is obvious that these data do not allow to constrain the value of $C^{\Lambda\Lambda,\Lambda\Lambda}_{1S0}$ more quantitatively. Note that in Figs.~\ref{fig:4.1} (a) and (c) one clearly sees the opening of the inelastic $\Xi^0n$ and $\Sigma^0\Lambda$ channels, respectively. 

The range of the $\Lambda\Lambda$ scattering length corresponding to the variations in $C^{\Lambda\Lambda,\Lambda\Lambda}_{1S0}$ is $a^{\Lambda\Lambda}_{1S0}=-1.38,...,-1.83$ fm. It is interesting to compare those values with the ones of the Nijmegen ESC04 model \cite{Rijken:2006kg} and the $YN$ model of Fujiwara et al. \cite{Fujiwara:2006yh}. These are the only $S=-2$ baryon-baryon interactions for which a direct comparison of the binding energy of the Nagara event, i.e. of the ${}_{\Lambda\Lambda}^{\;\;\;6}{\rm He}$ hypernucleus, with corresponding predictions based on those models is available. Specifically, Fujiwara et al. used their $\Xi N$ and $YY$ interactions in a Faddeev calculation where they considered the ${}_{\Lambda\Lambda}^{\;\;\;6}{\rm He}$ nucleus as an $\alpha$-particle and two baryons having strangeness $S=-2$. The two-$\Lambda$ separation energy, defined as $\Delta B_{\Lambda\Lambda}=B_{\Lambda\Lambda}({}^{\;\;\;6}_{\Lambda\Lambda}{\rm He})-2B_{\Lambda}({}^{5}_{\Lambda}{\rm He})$, obtained for their model is close to the experimental number of $1.01\pm0.20$ MeV \cite{Takahashi:2001nm}. The recent Nijmegen ESC04 model also reproduces the two-$\Lambda$ separation energy correctly \cite{Rijken:2006kg}.

The $\Lambda\Lambda$ scattering length given for the Nijmegen ESC04d interaction is $a^{\Lambda\Lambda}_{1S0}=-1.32$ fm \cite{Rijken:2006ep}. The value for the $\Lambda\Lambda$ interaction based on the constituent quark model of Fujiwara et al. is $a^{\Lambda\Lambda}_{1S0}=-0.81$ fm, see \cite{Fujiwara:2006yh}. The range of values predicted by the chiral EFT is close to the one of the Nijmegen models. Although the Nijmegen and the quark model have different $\Lambda\Lambda$ scattering lengths, both give a good reproduction of the experimentally observed two-$\Lambda$ separation energy, as mentioned. Obviously, from the $\Lambda\Lambda$ scattering length alone one can not draw any conclusions on the magnitude of the two-$\Lambda$ separation energy. Thus, the only reliable way to determine the two-$\Lambda$ separation energy corresponding to the chiral EFT interaction consists in a concrete calculation of doubly-strange hypernuclei. This has not been done so far. Clearly, such a calculation might help to further constrain the size of $C^{\Lambda\Lambda,\Lambda\Lambda}_{1S0}$.

\begin{figure}[t]
\resizebox{\textwidth}{!}{%
  \includegraphics*[2.0cm,17.0cm][15.65cm,26.8cm]{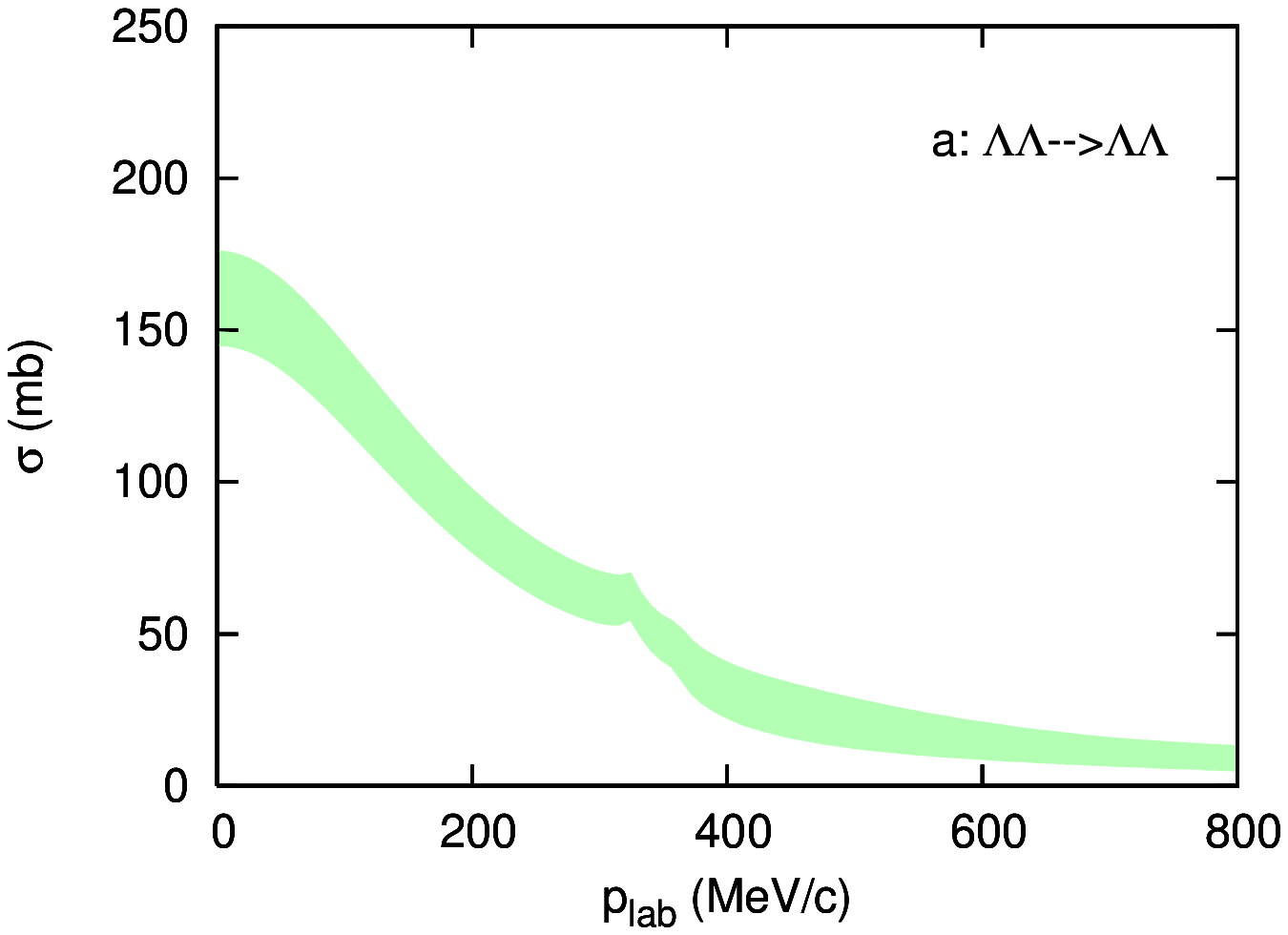}
  \includegraphics*[2.0cm,17.0cm][15.65cm,26.8cm]{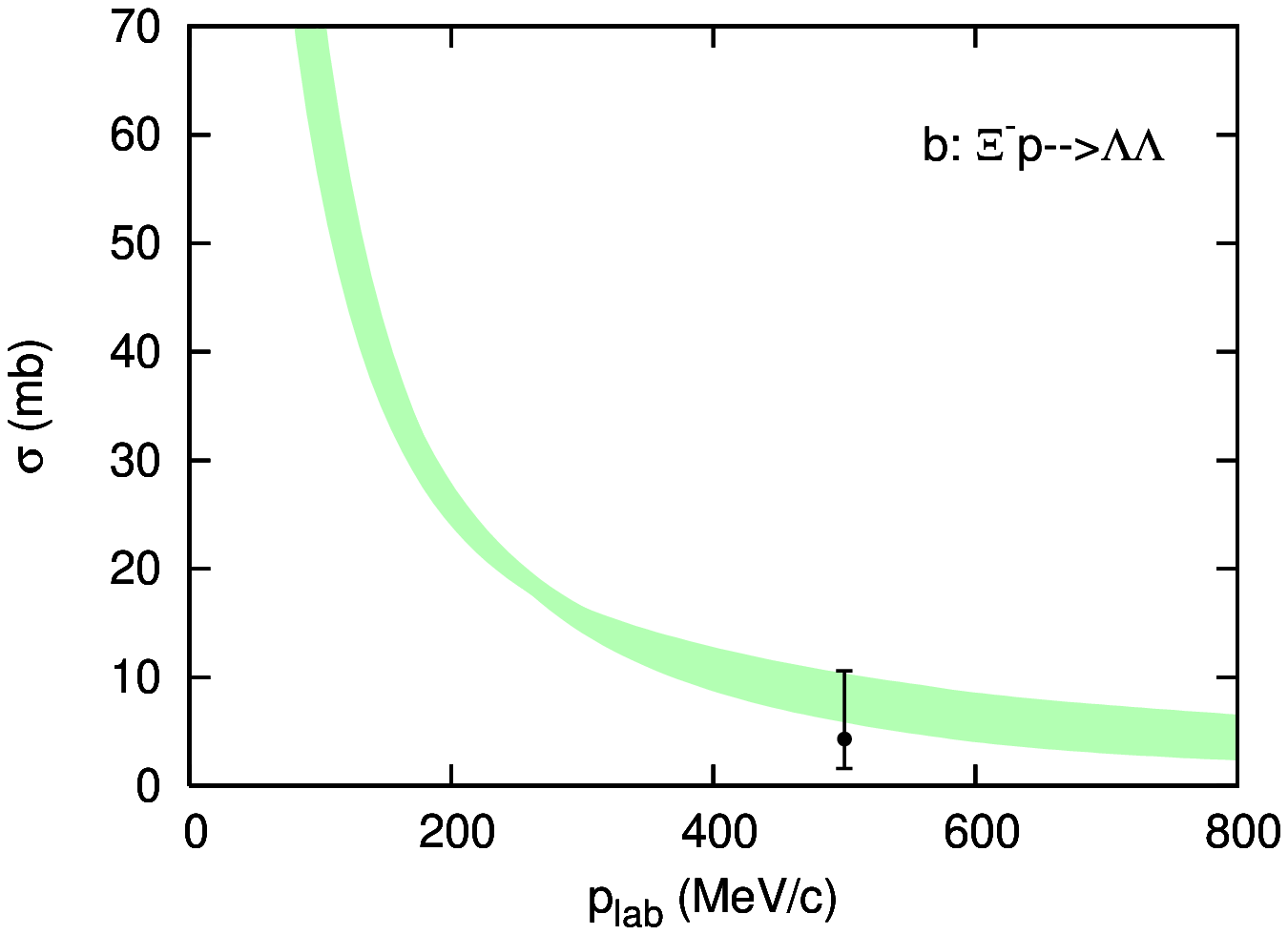}
}
\hfill \break
\resizebox{\textwidth}{!}{%
  \includegraphics*[2.0cm,17.0cm][15.65cm,26.8cm]{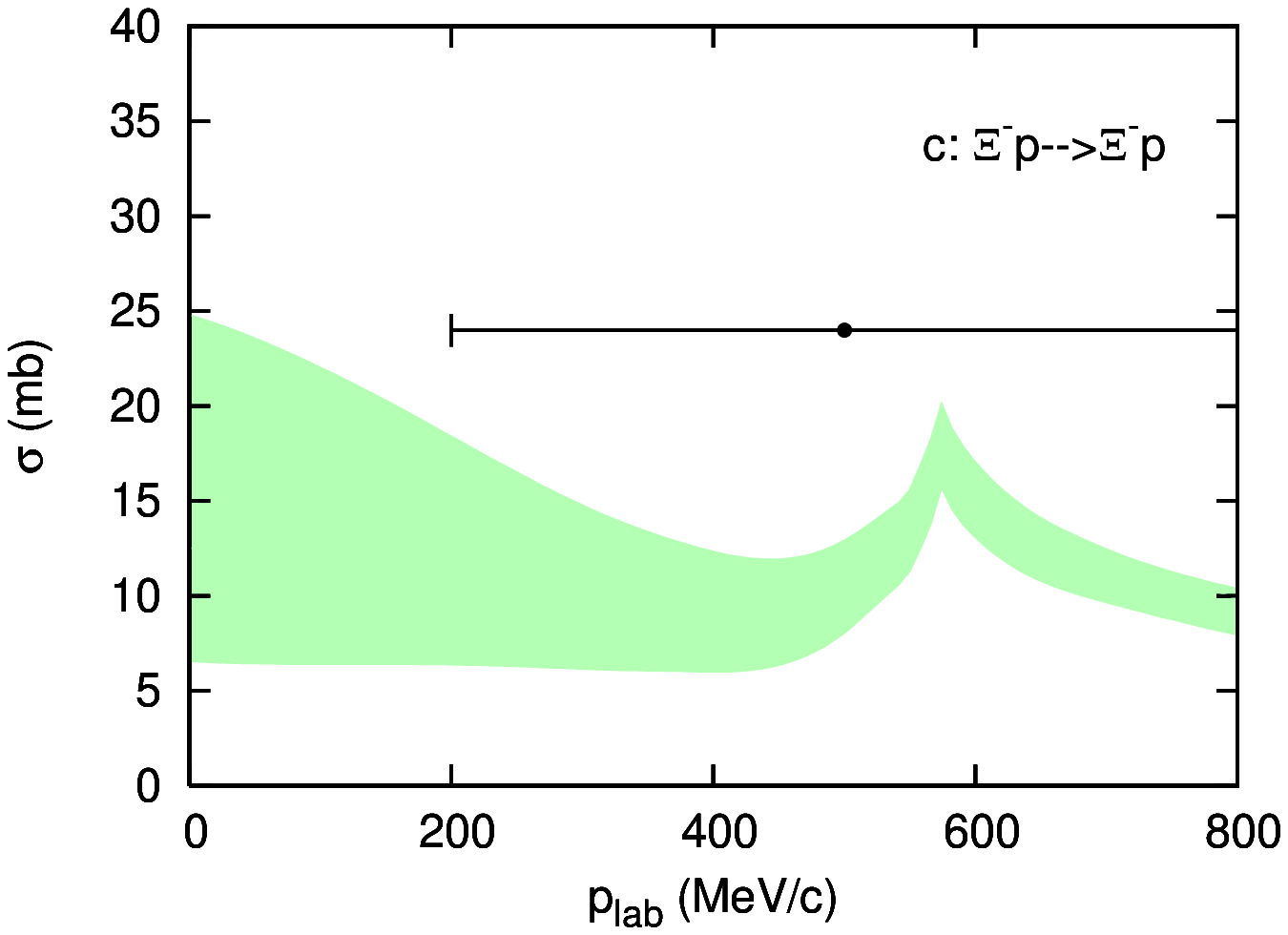}
  \includegraphics*[2.0cm,17.0cm][15.65cm,26.8cm]{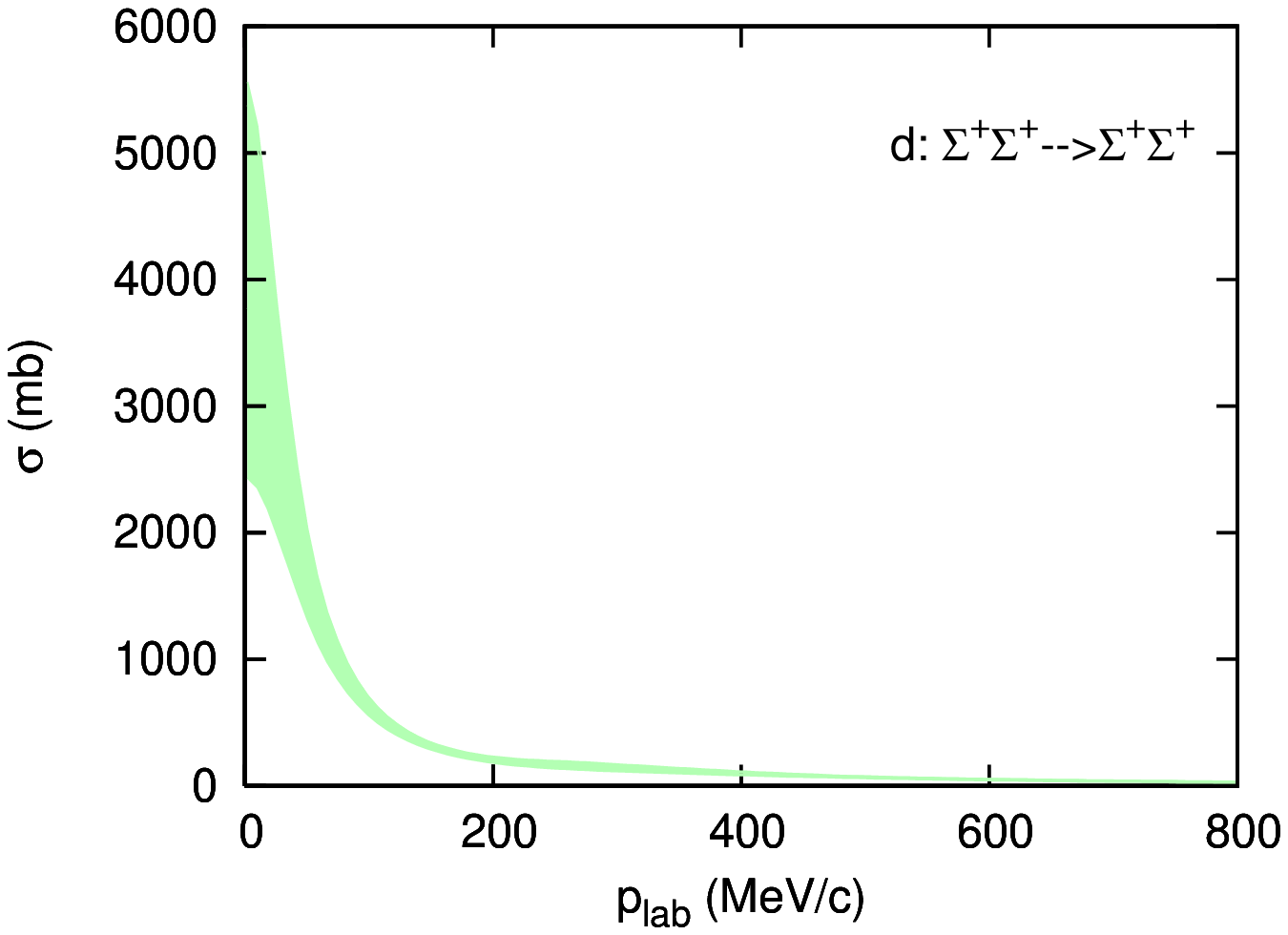}
}
\hfill \break
\resizebox{\textwidth}{!}{%
  \includegraphics*[2.0cm,17.0cm][15.65cm,26.8cm]{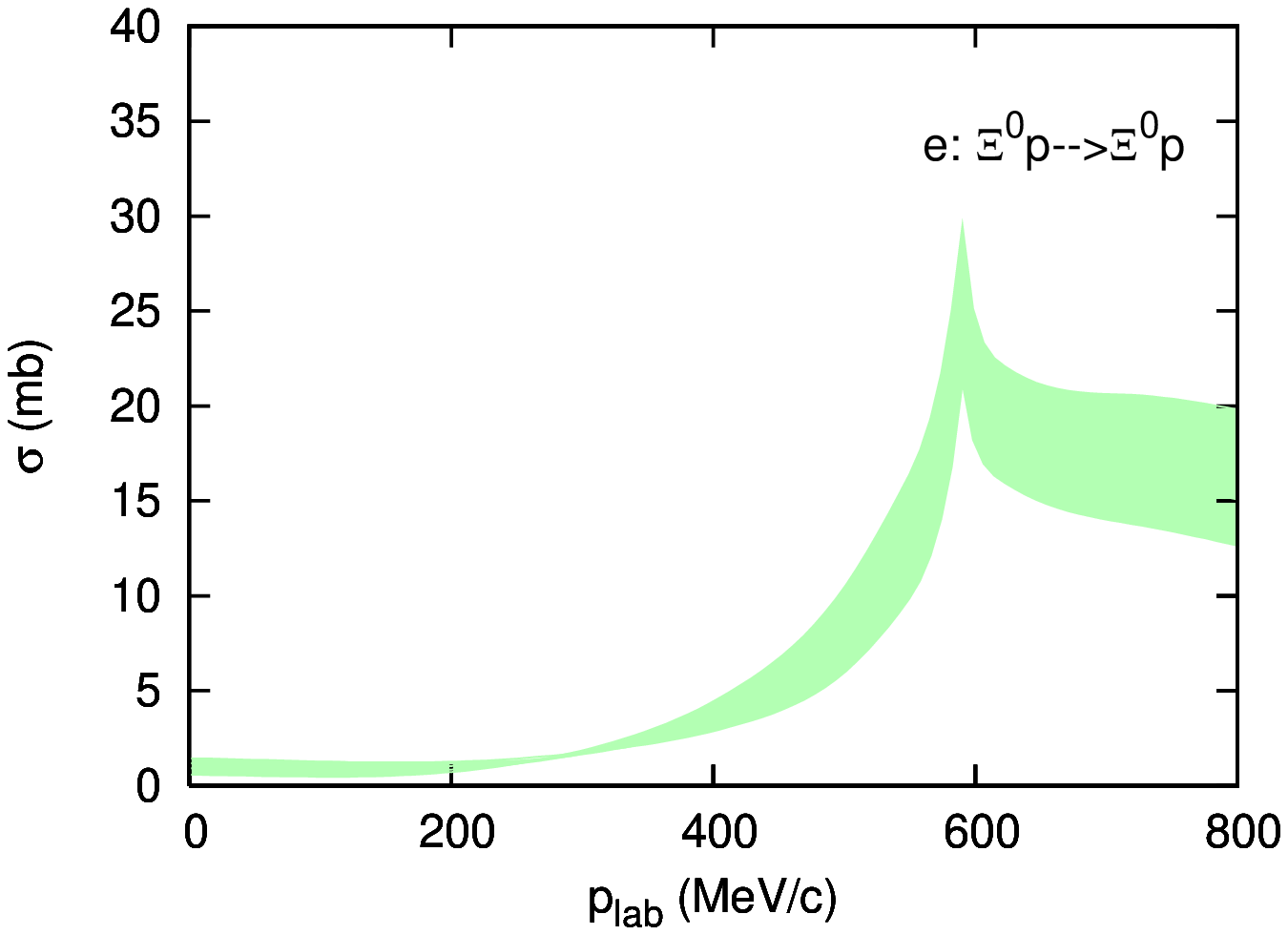}
  \includegraphics*[2.0cm,17.0cm][15.65cm,26.8cm]{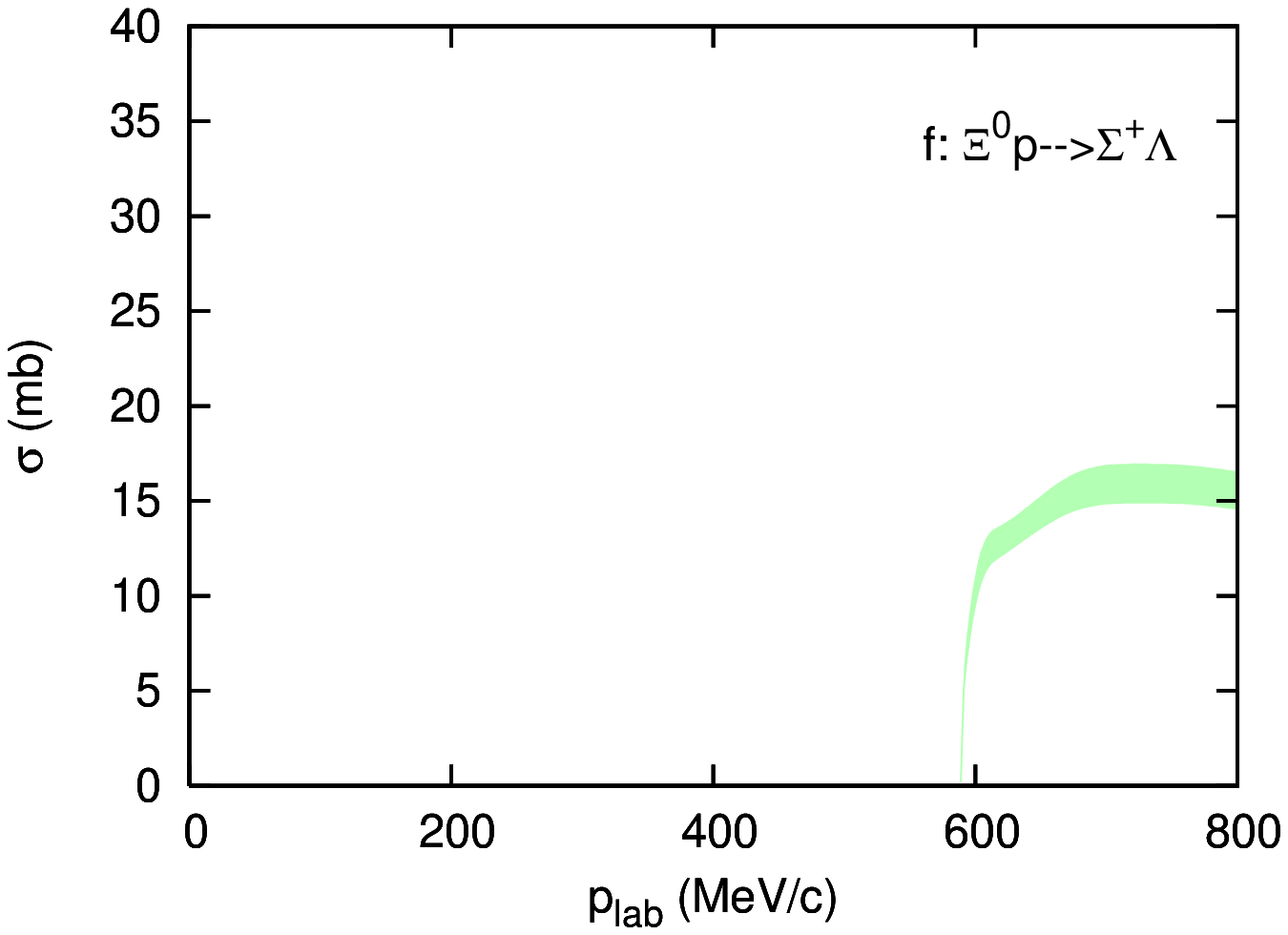}
}
%\vspace{0.005cm}
\caption{Total cross sections for the reactions $\Lambda\Lambda \to \Lambda\Lambda$, $\Xi^-p \to \Lambda\Lambda$, $\Xi^-p \to \Xi^-p$, $\Sigma^+\Sigma^+ \to \Sigma^+\Sigma^+$, $\Xi^0p \to \Xi^0p$, and $\Xi^0p \to \Sigma^+\Lambda$, as a function of $p_{\rm lab}$. The shaded band shows the chiral EFT results for variations of the cut-off in the range $\Lambda$=550,...,700 MeV. The results in (a)-(c) depend on the additional contact term $C^{\Lambda\Lambda,\Lambda\Lambda}_{1S0}$ while the other cross sections are genuine predictions. The experimental cross sections in (b) and (c) are taken from Ref. \cite{Ahn:2005jz}.
}
\label{fig:4.2}
\end{figure}

In order to study the cut-off dependence of the chiral EFT predictions, we first perform a reference calculation with $C^{\Lambda\Lambda,\Lambda\Lambda}_{1S0}=$0 and with the cut-off value of $\Lambda$=600 MeV. In subsequent calculations for other cut-off values we then vary $C^{\Lambda\Lambda,\Lambda\Lambda}_{1S0}$ in such a way that the $\Lambda\Lambda$ scattering length remains practically unchanged. We considered cut-off values in the range of 550,...,700 MeV like we did in Ref. \cite{Polinder:2006zh}. This range is also similar to the one considered in the $NN$ case, see, e.g. Refs.~\cite{Epelbaum:2002ji,Epelbaum:2003xx}. Results for the $\Lambda\Lambda$, $\Xi^0p$, and $\Sigma^+\Sigma^+$ scattering lengths are listed in Table~\ref{tab:4.1} together with the values for the additional contact term.
\begin{table}[t]
\caption{The $\Lambda\Lambda$, $\Xi^0p$, $\Sigma^+\Sigma^+$ singlet and $\Xi^0p$ triplet scattering lengths and effective ranges (in fm) for various cut-off values (in MeV). The last row shows the values for the additional contact term (in $10^4 {\rm GeV}^{-2}$) being relevant for the $\Lambda\Lambda$ channel.}
\label{tab:4.1}
\vspace{0.2cm}
\centering
\begin{tabular}{|l|rrrr|}
\hline
$\Lambda$ &$550$ &$600$ &$650$ &$700$\\
\hline
$a^{\Lambda\Lambda}_{1S0}$ &$-1.52$ &$-1.52$ &$-1.54$ &$-1.67$ \\
$r^{\Lambda\Lambda}_{1S0}$ &$0.82$  &$0.59$  &$0.31$  &$0.34$  \\
%&&&& \\
\hline
$a^{\Xi^0p}_{1S0}$ &$0.21$  &$0.19$        &$0.17$  &$0.13$ \\
$r^{\Xi^0p}_{1S0}$ &$-30.7$ &$-37.7$       &$-52.8$ &$-98.5$ \\
$a^{\Xi^0p}_{3S1}$ &$0.02$  &$0.00$        &$0.02$  &$0.03$ \\
$r^{\Xi^0p}_{3S1}$ &$968$   &$>\!\!10^4$   &$1166$  &$548$ \\
%&&&& \\
\hline
$a^{\Sigma^+\Sigma^+}_{1S0}$ &$-6.23$ &$-7.76$  &$-9.42$ &$-9.27$ \\
$r^{\Sigma^+\Sigma^+}_{1S0}$ &$2.17$  &$2.00$   &$1.88$  &$1.88$ \\
%&&&& \\
\hline
$C^{\Lambda\Lambda,\Lambda\Lambda}_{1S0}$  & $-0.0165$ & $0.0000$ 
& $0.0578$ & $0.0598$ \\
\hline
\end{tabular}
\end{table}
Cross sections for those reactions and some more $S=-2$ channels 
are presented in Fig.~\ref{fig:4.2}. 
The shaded bands show the cut-off dependence. We remark that the cross sections in Fig.~\ref{fig:4.2} (a)-(c) depend on the additional contact term, whereas the cross sections in Fig.~\ref{fig:4.2} (d)-(f) are independent of this term. Thus, the latter results are genuine predictions of the $YN$ EFT interaction \cite{Polinder:2006zh}. Obviously, the chiral EFT results remain consistent with the experimental cross sections for the considered range of variations of the cut-off, cf. Figs.~\ref{fig:4.2} (b) and (c). In Figs.~\ref{fig:4.2} (a), (c), and (e) one clearly sees the opening of the inelastic $\Xi^0n$, $\Sigma^0\Lambda$ and $\Sigma^+\Lambda$ channels, respectively. Fig.~\ref{fig:4.2} (e) also reveals that the pure $I=1$ component of the $\Xi N$ cross section is predicted to be quite small at low and intermediate energies, but it increases rapidly near the opening of the $\Sigma^+\Lambda$ channel. This behavior is very similar to that found for the Nijmegen NSC97f model \cite{Stoks:1999bz} and the quark model of Fujiwara et al. \cite{Fujiwara:2006yh}. From Fig.~\ref{fig:4.2} (d) one observes that the $\Sigma^+\Sigma^+$ cross section is rather large near threshold. This is also reflected in the corresponding scattering length, which is large too (cf. Table \ref{tab:4.1}) and, therefore, rather sensitive to cut-off variations. We should mention, however, that the results in this channel will change significantly once the Coulomb interaction is taken into account. Indeed, exploratory calculations including the Coulomb force showed that the scattering lengths are reduced by about 40\% and the sensitivity to the cut-off becomes much weaker. 
%%%%%%%%%%%%%%%%%%%%%%%%%%%%%%%%%%%%%%%%%%%%%%%%%%%%%%%%%%%%%%%%%%%%%%%%%%%%%%%

\section{Summary and outlook}
\label{chap:5}
In this letter we have presented first results for the doubly strange $\Xi N$ and $YY$ interactions ($Y = \Lambda, \Sigma$) obtained within a chiral effective field theory approach based on the Weinberg power counting, derived analogous to the $YN$ system studied in \cite{Polinder:2006zh}, by relating the $S=-2$ baryon-baryon interactions via ${\rm SU(3)}_{\rm f}$ symmetry to the $YN$ interactions.

The LO chiral potential consists of two pieces: firstly, the longer-ranged one-pseudo{\-}scalar-meson exchanges and secondly, shorter ranged four-baryon contact terms without derivatives.  In addition to the five contact terms, already present and fixed in our study of the $YN$ interaction, there appears an additional sixth contact term in the $\Xi N$ and $YY$ systems that can only be fixed in the $S=-2$ sector. The reaction amplitude is obtained by solving a regularized coupled-channels Lippmann-Schwinger equation for the LO chiral potential. 
We used an exponential regulator function to regularize the potential and applied cut--offs in the range between $550$ and $700$ MeV. In order to incorporate the correct physical thresholds, we solved the Lippmann-Schwinger equation in the particle basis.

We showed that the chiral EFT predictions are consistent with the recently deduced doubly strange scattering cross sections. Furthermore, a moderately attractive $\Lambda\Lambda$ interaction could be achieved for natural values of the additional contact term -- in line with recent empirical information on doubly strange hypernuclei. The presently available scattering data are, however, not sufficient to determine the value of the additional contact term more quantitatively.

It is expected that in the coming years better-quality data on the fundamental $\Xi N$ and $YY$ interactions as well as much more information about the physics of hypernuclei will become available at the new facilities J-PARC (Tokai, Japan) and FAIR (Darmstadt, Germany). The chiral EFT developed here can then be used to analyze these upcoming data in a model-independent way.

%%%%%%%%%%%%%%%%%%%%%%%%%%%%%%%%%%%%%%%%%%%%%%%%%%%%%%%%%%%%%%%%%%%%%%%%%%%%%%%
\ack
This research is part of the EU Integrated Infrastructure Initiative Hadron Physics Project under contract number RII3-CT-2004-506078. Work supported in part by DFG (SFB/TR 16, ``Subnuclear Structure of Matter'') and by the EU Contract No.~MRTN-CT-2006-035482, ``FLAVIAnet''.
%\bibliographystyle{elsart-num}
%\bibliography{yyeft}
%\end{document}

\end{document}